\documentclass[final,3p,times]{elsarticle}
\usepackage{amssymb}
\usepackage[fleqn]{amsmath}
\usepackage{amsthm}
\usepackage{subfigure}
\usepackage{graphicx}

\biboptions{numbers,sort&compress}
\usepackage{hyperref}
\journal{journal}

\begin{document}

\begin{frontmatter}

\title{Bright-dark mixed $N$-soliton solutions of the multi-component Mel'nikov system}

\author{Zhong Han$^{{\rm a,b}}$}

\author{Yong Chen$^{{\rm a,b}}$ \corref{cor1} }

\ead{ychen@sei.ecnu.edu.cn}
\cortext[cor1]{Corresponding author.}

\address{$^{{\rm a}}$Shanghai Key Laboratory of Trustworthy Computing, East China Normal University, Shanghai, 200062, People's Republic of China}
\address{$^{{\rm b}}$MOE International Joint Lab of Trustworthy Software, East China Normal University, Shanghai, 200062, People's Republic of China}

\begin{abstract}
By virtue of the KP hierarchy reduction technique, we construct the general bright-dark mixed $N$-soliton solution to the multi-component Mel'nikov system comprised of multiple (say $M$) short-wave components and one long-wave component with all possible combinations of nonlinearities including all-positive, all-negative and mixed types. Firstly, the two-bright-one-dark (2-b-1-d) and one-bright-two-dark (1-b-2-d) mixed $N$-soliton solutions in short-wave components of the three-component Mel'nikov system are derived in detail. Then we extend our analysis to the $M$-component Mel'nikov system to obtain its general mixed $N$-soliton solution. The formula obtained unifies the all-bright, all-dark and bright-dark mixed $N$-soliton solutions. For the collision of two solitons, the asymptotic analysis shows that for a $M$-component Mel'nikov system with $M \geq 3$,
inelastic collision takes place, resulting in energy exchange among the short-wave components supporting bright solitons only if the bright solitons appear at least in two short-wave components. Whereas, the dark solitons in the short-wave components and the bright solitons in the long-wave component always undergo elastic collision which just accompanied by a position shift.
\end{abstract}

\end{frontmatter}

\section{Introduction}

The study of multi-component nonlinear systems is of great interest as the interaction of multiple waves may result in some new physical phenomenons \cite{pre2,pre3,pre1,pre4}. Of particular interest is the multi-component generalization of the nonlinear Schr\"{o}dinger (NLS) equation \cite{gener3,gener4,gener2,feng,yan,ling1}, which has been considered as the generic model to describe the evolution of slowly varying wave packets in nonlinear wave systems.
What's more, the study on nonlinear systems describing the interaction of long waves with short wave packets in nonlinear dispersive media has received much attentions in recent years \cite{lak1,lak2,lak3,liu,chen1,chen2}. Such systems have found wide applications in the fields of hydrodynamics, nonlinear optics, plasma physics and so on \cite{yoko1,yoko2,yoko3,xiaoen}.

It is desirable to extend the studies to multi-component cases since a variety of complex systems such as nonlinear optical
fibres \cite{pre1}, Bose-Einstein condensates \cite{bose} etc usually involve more than one component. In the real physical systems, the nonlinearities can be positive or negative, depending on the physical situations \cite{segure}. For instance, in Bose-Einstein condensates \cite{bose}, the nonlinear coefficients take positive or negative when the
interaction between the atoms is repulsive or attractive.
For some multi-component systems \cite{lak1,lak3,chen2}, it has been found that the solitons exhibit certain energy-exchanging inelastic collision behaviors, which have not been found in the single-component counterparts and may be used to realize multi-state logic and soliton collision-based computing.
In the current paper, we consider the multi-component generalization of the Mel'nikov system \cite{mel1,mel2,mel3,mel4}
\begin{align}
& \textmd{i}\Phi_y=\Phi_{xx}+u \Phi,\label{j1}\\
&u_{xt}+u_{xxxx}+3(u^2)_{xx}-3 u_{yy}+\sigma (\Phi{\Phi}^\ast )_{xx}=0,\label{j2}
\end{align}
where $\sigma=\pm 1$, $u$ is the real long-wave amplitude and $\Phi$ is the complex short-wave amplitude; the asterisk means complex conjugate hereafter and the subscripts $t$ and $x,y$ denote partial differentiation with respect to time and space, respectively. This system is introduced by Mel'nikov and can be used to describe the interaction of long waves with short wave packets propagating on the $x$-$y$ plane at an angle to each other. The Mel'nikov system (\ref{j1})-(\ref{j2}) can be extended to a multi-component case
\begin{align}
& \textmd{i}\Phi^{(k)}_y=\Phi^{(k)}_{xx}+u \Phi^{(k)},\ \ \ \ \ \ k=1,2,\cdots, M,\label{j3}\\
&u_{xt}+u_{xxxx}+3(u^2)_{xx}-3 u_{yy}+\Big(\sum^M_{k=1} \sigma_k\Phi^{(k)}{\Phi^{(k)}}^\ast \Big)_{xx}=0,\ \ \ \ \ \ \sigma_k=\pm 1,\label{j4}
\end{align}
which describes the interaction of a long wave $u$ with multiple (say $M$) short wave packets $\Phi^{(k)}$.
The system (\ref{j3})-(\ref{j4}) is refered as the $M$-component Mel'nikov system hereafter.

The Mel'nikov system (\ref{j1})-(\ref{j2}) admits boomeron type solutions \cite{mel3}, and its multi-soliton solution is obtained in Ref.\cite{mel4} through the matrices theory.
Its bright- and dark-types soliton solutions have been derived from the Wronskian solutions of the KP hierarchy equaions \cite{hase}.
The Painlev\'{e} analysis and exponentially localized dromion type solutions of this system are reported in Ref.\cite{india}.
In a recent work, its rogue wave solution is obtained by using the Hirota's bilinear method \cite{qin}.
However, to the best of our knowledge, the general bright-dark mixed $N$-soliton solution of the multi-component Mel'nikov system (\ref{j3})-(\ref{j4}) has not been reported so far.

The goal of the present paper is to construct the general bright-dark mixed $N$-soliton solution to the multi-component Mel'nikov system (\ref{j3})-(\ref{j4}) with all possible combinations of nonlinearities including all-positive, all-negative and mixed types through the KP hierarchy reduction technique.
Moreover, the dynamics of single and two solitons are also discussed in detail.
For the collision of two solitons in a $M$-component Mel'nikov system with $M \geq 3$, it can be shown that inelastic collision takes place, which results in energy exchange
among the bright parts of the mixed solitons in short-wave components only if the bright parts appear at least in two short-wave components. While the dark parts of the mixed solitons in short-wave components and the bright solitons in long-wave component always undergo standard elastic collision.

It is worth noting that the KP hierarchy reduction technique to derive soliton solutions of integrable systems is an effective and elegant method, which is
firstly developed by the Kyoto school \cite{jimbo1} in the 1970s. This method has been applied to get soliton solutions of the NLS equation, the modified KdV equation and the Davey-Stewartson (DS) equation. Additionally, the pseudo-reduction of the two-dimensional Toda lattice hierarchy to constrained KP systems with dark soliton solutions is introduced in Ref.\cite{will1}, and the reduction to constrained KP systems with bright soliton solutions from multi-component KP hierarchy is established in Ref.\cite{will2}. Based on this method, Ohta et al \cite{ohta} construct the general $N$-dark-dark soliton solutions for a two-coupled NLS equations (Manakov system). Also using this method, the general bright-dark mixed $N$-soliton solution of the vector NLS equations is investigated by Feng \cite{feng}. Most recently, this method is used to obtain the $N$-dark soliton \cite{chen1} and bright-dark mixed $N$-soliton \cite{chen2} solutions of the multi-component Yajima-Oikawa (YO) system. In some other recent works, the KP hierarchy reduction technique has also been applied to derive rogue wave solutions of integrable systems \cite{ohta2,ohta3,ohta4}, see also the literatures \cite{chen3,shi,he}.

This paper is organized as below. In section 2, the general two-bright-one-dark and one-bright-two-dark mixed $N$-soliton solutions in Gram determinant form of the three-component Mel'nikov system are derived in detail. Besides, the dynamics of single and two solitons are also discussed. Section 3 devotes to extend the similar analysis to obtain the general $m$-bright-($M-m$)-dark mixed $N$-soliton solution of the $M$-component Mel'nikov system. The last section is allotted for conclusion.

\section{Bright-dark Mixed $N$-Soliton Solution of the three-component Mel'nikov System}
We first consider the general bright-dark mixed $N$-soliton solution to the three-component Mel'nikov system
\begin{align}
& \textmd{i}\Phi^{(k)}_y=\Phi^{(k)}_{xx}+u \Phi^{(k)},\ \ \ \ k=1,2,3,\label{j5}\\
&u_{xt}+u_{xxxx}+3(u^2)_{xx}-3 u_{yy}+\Big(\sum^3_{k=1} \sigma_k\Phi^{(k)}{\Phi^{(k)}}^\ast \Big)_{xx}=0,\label{j8}
\end{align}
where $\sigma_k=\pm 1$ for $k=1,2,3$. For the three-component Mel'nikov system, the mixed-type vector solitons in the short-wave components consist of two types: two-bright-one-dark (2-b-1-d) and one-bright-two-dark (1-b-2-d). These two types of soliton solutions will be derived in the subsequent two subsections, respectively.

\subsection{2-b-1-d mixed soliton solution}

Without loss of generality, assuming the $\Phi^{(1)}$ and $\Phi^{(2)}$ components are of bright type while the $\Phi^{(3)}$ component is of dark type. We introduce the dependent variable transformations
\begin{align}\label{jj7}
& \Phi^{(1)}= \frac{g^{(1)}}{f},\ \ \ \ \Phi^{(2)}= \frac{g^{(2)}}{f},\ \ \ \ \Phi^{(3)}=\rho_1 {\rm e}^{\textmd{i}\theta_1} \frac{h^{(1)}}{f},\ \ \ \ u=2 (\log f)_{xx},
\end{align}
where $g^{(1)}, g^{(2)}$ and $h^{(1)}$ are complex functions; $f$ is a real function; $\theta_1=\alpha_1x+\alpha^2_1y+\beta_1(t)$, $\alpha_1$ and $\rho_1$ are real constants, $\beta_1(t)$ is a real function. Then the three-component Mel'nikov system (\ref{j5})-(\ref{j8}) is converted into the bilinear form
\begin{align}
& (D^2_x-{\rm i} D_y)g^{(k)} \cdot f=0,\ \ \ \ k=1,2,\label{j10}\\
& (D^2_x+2 \textmd{i} \alpha_1 D_x-{\rm i} D_y)h^{(1)} \cdot f=0,\label{j11}\\
& (D^4_x+D_xD_t-3D^2_y)f\cdot f=-\sum^2_{k=1}\sigma_k g^{(k)}{g^{(k)}}^\ast+\sigma_3 \rho^2_1 (f^2-h^{(1)}{h^{(1)}}^\ast),\label{j13}
\end{align}
where the Hirota's bilinear operator $D$ is defined as
\begin{align}
& D^l_xD^m_yD^n_tf(x,y,t)\cdot g(x,y,t)=\Big(\frac{\partial}{\partial x}-\frac{\partial}{\partial x'}\Big)^l\Big(\frac{\partial}{\partial y}-\frac{\partial}{\partial y'}\Big)^m\Big(\frac{\partial}{\partial t}-\frac{\partial}{\partial t'}\Big)^nf(x,y,t)\cdot g(x',y',t')|_{x=x',y=y',t=t'}.
\end{align}

In what follows, we proceed to show how the mixed $N$-soliton solution is derived through the KP hierarchy reduction technique. To this end, we consider a three-component KP hierarchy with one copy of shifted singular point ($c_1$)
\begin{align}
& (D^2_{x_1}-D_{x_2})\tau_{1,0}(k_1) \cdot \tau_{0,0}(k_1)=0,\label{j20}\\
& (D^2_{x_1}-D_{x_2})\tau_{0,1}(k_1) \cdot \tau_{0,0}(k_1)=0,\label{j21}\\
& (D^2_{x_1}-D_{x_2}+2c_1D_{x_1})\tau_{0,0}(k_1+1) \cdot \tau_{0,0}(k_1)=0,\label{j22}\\
& (D^4_{x_1}-4D_{x_1}D_{x_3}+3D^2_{x_2})\tau_{0,0}(k_1)\cdot \tau_{0,0}(k_1)=0,\label{j23}\\
& D_{x_1}D_{y^{(1)}_1}\tau_{0,0}(k_1) \cdot \tau_{0,0}(k_1)=-2\tau_{1,0}(k_1)  \tau_{-1,0}(k_1),\label{j24}\\
& D_{x_1}D_{y^{(2)}_1}\tau_{0,0}(k_1) \cdot \tau_{0,0}(k_1)=-2\tau_{0,1}(k_1)  \tau_{0,-1}(k_1),\label{j25}\\
& (D_{x_1}D_{x^{(1)}_{-1}}-2)\tau_{0,0}(k_1) \cdot \tau_{0,0}(k_1)=-2\tau_{0,0}(k_1+1)  \tau_{0,0}(k_1-1).\label{j25b}
\end{align}
Based on the Sato theory for KP hierarchy \cite{jimbo1}, the bilinear equations (\ref{j20})-(\ref{j25b}) have the Gram determinant tau function solution
\begin{align}
& \tau_{0,0}(k_1)=\left| \begin{array}{ccccc}
A & I  \\
-I & B
\end{array} \right|,\\
& \tau_{1,0}(k_1)=\left| \begin{array}{ccccc}
A & I  & \Omega^{\textmd{T}}\\
-I & B & \mathbf{0}^{\textmd{T}}\\
\mathbf{0} & -\bar{\Psi} & 0
\end{array} \right|,\ \ \ \ \ \
 \tau_{-1,0}(k_1)=\left| \begin{array}{ccccc}
A & I  & \mathbf{0}^{\textmd{T}}\\
-I & B & \Psi^{\textmd{T}}\\
-\bar{\Omega} & \mathbf{0} & 0
\end{array} \right|,\\
& \tau_{0,1}(k_1)=\left| \begin{array}{ccccc}
A & I  & \Omega^{\textmd{T}}\\
-I & B & \mathbf{0}^{\textmd{T}}\\
\mathbf{0} & -\bar{\Upsilon} & 0
\end{array} \right|,\ \ \ \ \ \ \
 \tau_{0,-1}(k_1)=\left| \begin{array}{ccccc}
A & I  & \mathbf{0}^{\textmd{T}}\\
-I & B & \Upsilon^{\textmd{T}}\\
-\bar{\Omega} & \mathbf{0} & 0
\end{array} \right|,
\end{align}
where $\mathbf{0}$ is an $N$-component zero-row vector; $I$ is an $N \times N$ identity matrix; $A$ and $B$ are $N \times N$ matrices whose elements are defined respectively as
\begin{align*}
& a_{ij}(k_1)=\frac{1}{p_i+\bar{p}_j} \Big(-\frac{p_i-c_1}{\bar{p}_j+c_1}\Big)^{k_1} \textmd{e}^{\xi_i+\bar{\xi}_j},\ \ \ \ \ \ b_{ij}=\frac{1}{q_i+\bar{q}_j}\textmd{e}^{\eta_i+\bar{\eta}_j}+ \frac{1}{r_i+\bar{r}_j}\textmd{e}^{\chi_i+\bar{\chi}_j},
\end{align*}
meanwhile, $\Omega, \Psi, \Upsilon, \bar{\Omega}, \bar{\Psi}$ and $\bar{\Upsilon}$ are $N$-component row vectors
\begin{align*}
&  \Omega=(\textmd{e}^{\xi_1},\textmd{e}^{\xi_2},\cdots,\textmd{e}^{\xi_N}),\ \ \ \ \ \ \Psi=(\textmd{e}^{\eta_1},\textmd{e}^{\eta_2},\cdots,\textmd{e}^{\eta_N}),\ \ \ \ \ \ \Upsilon=(\textmd{e}^{\chi_1},\textmd{e}^{\chi_2},\cdots,\textmd{e}^{\chi_N}),\\
&  \bar{\Omega}=(\textmd{e}^{\bar{\xi}_1},\textmd{e}^{\bar{\xi}_2},\cdots,\textmd{e}^{\bar{\xi}_N}),\ \ \ \ \ \ \bar{\Psi}=(\textmd{e}^{\bar{\eta}_1},\textmd{e}^{\bar{\eta}_2},\cdots,\textmd{e}^{\bar{\eta}_N}),\ \ \ \ \ \ \bar{\Upsilon}=(\textmd{e}^{\bar{\chi}_1},\textmd{e}^{\bar{\chi}_2},\cdots,\textmd{e}^{\bar{\chi}_N}),
\end{align*}
with
\begin{align*}
& \xi_i=\frac{1}{p_i-c_1}x^{(1)}_{-1}+p_ix_1+p^2_ix_2+p^3_ix_3+\xi_{i0},\ \ \ \ \ \  \bar{\xi}_j=\frac{1}{\bar{p}_j+c_1}x^{(1)}_{-1}+\bar{p}_jx_1-\bar{p}^2_jx_2+\bar{p}^3_jx_3+\bar{\xi}_{j0},\\
& \eta_i=q_iy^{(1)}_1+\eta_{i0},\ \ \ \ \ \  \bar{\eta}_j=\bar{q}_jy^{(1)}_1+\bar{\eta}_{j0},\ \ \ \ \ \
 \chi_i=r_iy^{(2)}_1+\chi_{i0},\ \ \ \ \ \  \bar{\chi}_j=\bar{r}_jy^{(2)}_1+\bar{\chi}_{j0},
\end{align*}
in which $p_i, \bar{p}_j, q_i, \bar{q}_j, r_i, \bar{r}_j, \xi_{i0}, \bar{\xi}_{j0}, \eta_{i0}, \bar{\eta}_{j0}, \chi_{i0}, \bar{\chi}_{j0}$ and $c_1$ are complex constants.

The proof of the bilinear equations (\ref{j20})-(\ref{j25b}) can be shown by means of the Grammian technique \cite{hirota,miyake}, which is omitted here.
We first consider complex conjugate reduction by assuming $x_1$, $x_3$, $x^{(1)}_{-1}$, $y^{(1)}_1$, $y^{(2)}_1$ are real; $x_2$ and $c_1$ are pure imaginary and by letting $p^*_j=\bar{p}_j,q^*_j=\bar{q}_j,r^*_j=\bar{r}_j,\xi^*_{j0}=\bar{\xi}_{j0},\eta^*_{j0}=\bar{\eta}_{j0}$ and $\chi^*_{j0}=\bar{\chi}_{j0}$, then it is easy to check that
\begin{align*}
a_{ij}(k_1)=a^*_{ji}(k_1),\ \ \ \ b_{ij}=b^*_{ji}.
\end{align*}
Furthermore, we define
\begin{align*}
f=\tau_{0,0}(0),\ \ \ \ g^{(1)}=\tau_{1,0}(0),\ \ \ \ g^{(2)}=\tau_{0,1}(0),\ \ \ \ h^{(1)}=\tau_{0,0}(1),
\end{align*}
hence, $f$ is real and
\begin{align*}
g^{(1)*}=-\tau_{-1,0}(0),\  \ \ \ g^{(2)*}=-\tau_{0,-1}(0),\  \ \ \ h^{(1)*}=\tau_{0,0}(-1),
\end{align*}
thus the bilinear equations (\ref{j20})-(\ref{j25b}) become to
\begin{align}
& (D^2_{x_1}-D_{x_2})g^{(k)} \cdot f=0,\ \ \ \ k=1,2,\label{j26}\\
& (D^2_{x_1}-D_{x_2}+2c_1D_{x_1})h^{(1)} \cdot f=0,\label{j27}\\
& (D^4_{x_1}-4D_{x_1}D_{x_3}+3D^2_{x_2})f\cdot f=0,\label{j28}\\
& D_{x_1}D_{y^{(k)}_1}f \cdot f=2g^{(k)}g^{(k)*},\ \ \ \ k=1,2,\label{j30}\\
& (D_{x_1}D_{x^{(1)}_{-1}}-2)f \cdot f=-2h^{(1)} h^{(1)*}\label{j31}.
\end{align}

Using the independent variable transformations
\begin{align}
& x_1=x,\ \ \ \ x_2=-\textmd{i}y,\ \ \ \ x_3=-8t,\label{j42}
\end{align}
i.e.,
\begin{align}
& \partial_x=\partial_{x_1},\ \ \ \ \partial_y=-\textmd{i}\partial_{x_2},\ \ \ \ \partial_{t}=-8\partial_{x_3},
\end{align}
equations (\ref{j26})-(\ref{j27}) become to equations (\ref{j10})-(\ref{j11}) by letting $c_1={\rm i}\alpha_1$.
Next, we show how to get equation (\ref{j13}) from equations (\ref{j28})-(\ref{j31}).

By row operations, $f$ can be rewritten as
\begin{align}
& f=\left| \begin{array}{ccccc}
A' & I  \\
-I & B'
\end{array} \right|,
\end{align}
where $A'$ and $B'$ are $N \times N$ matrices whose entries are
\begin{align*}
& a'_{ij}=\frac{1}{p_i+p^*_j},\ \ \ \ \ \
 b'_{ij}=\frac{1}{q_i+q^*_j}\textmd{e}^{\eta_i+\eta^*_j+\xi^*_i+\xi_j}+ \frac{1}{r_i+r^*_j}\textmd{e}^{\chi_i+\chi^*_j+\xi^*_i+\xi_j},
\end{align*}
with
\begin{align*}
& \eta_i+\xi^*_i=q_iy^{(1)}_1+\frac{1}{p^*_i+c_1}x^{(1)}_{-1}+p^*_ix_1-{p_i^*}^2x_2+{p_i^*}^3x_3+\xi^*_{i0}+\eta_{i0},\\
& \eta^*_j+\xi_j=q^*_jy^{(1)}_1+\frac{1}{p_j-c_1}x^{(1)}_{-1}+p_jx_1+p_j^2x_2+p_j^3x_3+\xi_{j0}+\eta^*_{j0},\\
& \chi_i+\xi^*_i=r_iy^{(2)}_1+\frac{1}{p^*_i+c_1}x^{(1)}_{-1}+p^*_ix_1-{p_i^*}^2x_2+{p_i^*}^3x_3+\xi^*_{i0}+\chi_{i0},\\
& \chi^*_j+\xi_j=r^*_jy^{(2)}_1+\frac{1}{p_j-c_1}x^{(1)}_{-1}+p_jx_1+p_j^2x_2+p_j^3x_3+\xi_{j0}+\chi^*_{j0}.
\end{align*}

Consider the following reduction conditions
\begin{align}
& 8{p_i^*}^3=\sigma_1q_i-\frac{\sigma_3\rho^2_1}{p_i^*+c_1},\ \ \ \ \ \ \  8p^3_j=\sigma_1q^*_j-\frac{\sigma_3\rho^2_1}{p_j-c_1},\\
& 8{p_i^*}^3=\sigma_2r_i-\frac{\sigma_3\rho^2_1}{p_i^*+c_1},\ \ \ \ \ \ \  8p^3_j=\sigma_2r^*_j-\frac{\sigma_3\rho^2_1}{p_j-c_1},
\end{align}
i.e.,
\begin{align}
& \frac{1}{q_i+q^*_j}=\frac{\sigma_1}{8({p_i^*}^3+p_j^3)+\frac{\sigma_3\rho^2_1(p_i^*+p_j)}{(p_i^*+\textmd{i}\alpha_1)(p_j-\textmd{i}\alpha_1)}},\\
& \frac{1}{r_i+r^*_j}=\frac{\sigma_2}{8({p_i^*}^3+p_j^3)+\frac{\sigma_3\rho^2_1(p_i^*+p_j)}{(p_i^*+\textmd{i}\alpha_1)(p_j-\textmd{i}\alpha_1)}},
\end{align}
we have the relation
\begin{align}
& 8\partial_{x_3}b'_{ij}=(\sigma_1\partial_{y^{(1)}_1}+\sigma_2\partial_{y^{(2)}_1}-\sigma_3\rho^2_1\partial_{x^{(1)}_{-1}})b'_{ij}.\label{n1}
\end{align}
Equation (\ref{n1}) immediately leads to
\begin{align}
& 8f_{x_3}=\sigma_1f_{y^{(1)}_1}+\sigma_2f_{y^{(2)}_1}-\sigma_3\rho^2_1f_{x^{(1)}_{-1}},\label{j38}
\end{align}
and the derivation of (\ref{j38}) with respect to $x_1$ reads
\begin{align}
& 8f_{x_1x_3}=\sigma_1f_{x_1y^{(1)}_1}+\sigma_2f_{x_1y^{(2)}_1}-\sigma_3\rho^2_1f_{x_1x^{(1)}_{-1}}.\label{j39}
\end{align}

On the other hand, equations (\ref{j30}) and (\ref{j31}) can be expanded as
\begin{align}
& f_{x_1y^{(1)}_1}f-f_{x_1}f_{y^{(1)}_{1}}=g^{(1)}g^{(1)*},\ \ \ \ \ \ \ \ \ f_{x_1y^{(2)}_1}f-f_{x_1}f_{y^{(2)}_{1}}=g^{(2)}g^{(2)*},\label{j40}
\end{align}
and
\begin{align}
& f_{x_1x^{(1)}_{-1}}f-f_{x_1}f_{x^{(1)}_{-1}}-f^2=-h^{(1)} h^{(1)*},\label{j41}
\end{align}
respectively. By using relations (\ref{j38}) and (\ref{j39}), from (\ref{j40}) and (\ref{j41}), we can arrive at
\begin{align}\label{j41n}
& -4D_{x_1}D_{x_3}f\cdot f=-\sigma_1g^{(1)}g^{(1)*}-\sigma_2g^{(2)}g^{(2)*}+\sigma_3\rho^2_1(f^2-h^{(1)}h^{(1)*}).
\end{align}
Also by the transformations (\ref{j42}),
from equations (\ref{j28}) and (\ref{j41n}), equation (\ref{j13}) is immediately obtained.

Under the variable transformations (\ref{j42}), the variables $y^{(1)}_1, y^{(2)}_1,x^{(1)}_{-1}$ become dummy variables, thus they can be treated as constants. Consequently, we can take $ {\rm e}^{\eta_{i}}=c^{(1)*}_i$, ${\rm e}^{\eta^*_{i}}=c^{(1)}_i$, $ {\rm e}^{\chi_{i}}=c^{(2)*}_i$, ${\rm e}^{\chi^*_{i}}=c^{(2)}_i$, $(i=1,2,\cdots,N)$ and define $ C_1=-(c^{(1)}_1,c^{(1)}_2,\cdots,c^{(1)}_N)$ and $ C_2=-(c^{(2)}_1,c^{(2)}_2,\cdots,c^{(2)}_N)$,
thus we have obtained the 2-b-1-d mixed $N$-soliton solution of the three-component Mel'nikov system (\ref{j5})-(\ref{j8})
\begin{align}\label{j44}
& f=\left| \begin{array}{ccccc}
A & I  \\
-I & B
\end{array} \right|,\ \ \ \ \ \ \
 g^{(k)}=\left| \begin{array}{ccccc}
A & I  & \Omega^{\textmd{T}}\\
-I & B & \mathbf{0}^{\textmd{T}}\\
\mathbf{0} & C_k & 0
\end{array} \right|,\ \ \ \ \ \ \
h^{(1)}=\left| \begin{array}{ccccc}
A^{(1)} & I  \\
-I & B
\end{array} \right|,
\end{align}
where the entries in $A,A^{(1)}$ and $B$ are defined as
\begin{align}\label{j45}
& a_{ij}=\frac{1}{p_i+p^*_j} \textmd{e}^{\xi_i+\xi^*_j},\ \ \ \ \ \ a^{(1)}_{ij}=\frac{1}{p_i+p^*_j} \Big(-\frac{p_i-\textmd{i}\alpha_1}{p^*_j+\textmd{i}\alpha_1}\Big) \textmd{e}^{\xi_i+\xi^*_j},\\ & b_{ij}=\Big(\sum_{k=1}^{2}\sigma_kc_i^{(k)*}c_j^{(k)}\Big)\Big[8({p_i^*}^3+p_j^3)+\frac{\sigma_3\rho^2_1(p_i^*+p_j)}{(p_i^*+\textmd{i}\alpha_1)(p_j-\textmd{i}\alpha_1)}\Big]^{-1},
\end{align}
respectively; $\Omega$ and $C_k$ are $N$-component row vectors
\begin{align}
& \Omega=(\textmd{e}^{\xi_1},\textmd{e}^{\xi_2},\cdots,\textmd{e}^{\xi_N}),\ \ \ \ \ \ C_k=-(c^{(k)}_1,c^{(k)}_2,\cdots,c^{(k)}_N),
\end{align}
with $ \xi_i=p_ix -\textmd{i}p^2_iy-8p^3_i t  + \xi_{i0}$, and $p_i$, $\xi_{i0}$, $c^{(k)}_i$, $(k=1,2;i=1,2,\cdots,N)$ are complex constants.

\subsubsection{One-soliton solution}
By taking $N=1$ in the formula (\ref{j44}), we can get the one-soliton solution. For this case, the tau functions can be rewritten as
\begin{align}
& f=1+E_{11^*}\textmd{e}^{\xi_1+\xi^*_1},\\
& g^{(k)}=c_1^{(k)}\textmd{e}^{\xi_1},\ \ \ \ \ k=1,2,\\
& h^{(1)}=1+F_{11^*}\textmd{e}^{\xi_1+\xi^*_1},
\end{align}
where
\begin{align*}
&E_{11^*}=\Big(\sum^2_{k=1}\sigma_kc_1^{(k)}c_1^{(k)*}\Big)\Big[8(p_1+p_1^*)(p_1^3+{p_1^*}^3)+\frac{\sigma_3\rho^2_1(p_1+p_1^*)^2}{(p_1^*+\textmd{i}\alpha_1)(p_1-\textmd{i}\alpha_1)}\Big]^{-1},\\
& F_{11^*}=-\frac{p_1-\textmd{i}\alpha_1}{p^*_1+\textmd{i}\alpha_1}E_{11^*},
\end{align*}
with
$ \xi_1=p_1x -\textmd{i}p^2_1y-8p^3_1 t  + \xi_{10}.$
Besides, the one-soliton solution is nonsingular only when $E_{11^*}>0$.

The 2-b-1-d mixed one-soliton solution can be expressed in the form
\begin{align}
& \Phi^{(k)}=\frac{c^{(k)}_1}{2}  \textmd{e}^{\textmd{i}\xi_{1I}-\eta_1} {\rm sech}(\xi_{1R}+\eta_1),\ \ \ \ \ k=1,2,\\
& \Phi^{(3)}=\frac{\rho_1}{2} \textmd{e}^{\textmd{i}\theta_{1}} [1+\textmd{e}^{2\textmd{i}\phi_1}+(\textmd{e}^{2\textmd{i}\phi_1}-1)\tanh(\xi_{1R}+\eta_1)],\\
& u=2p^2_{1R}{\rm sech}^2(\xi_{1R}+\eta_1),
\end{align}
where $\textmd{e}^{2 \eta_1}=E_{11^*}$, $\textmd{e}^{2\textmd{i}\phi_1}=-(p_1-\textmd{i}\alpha_1)/(p^*_1+\textmd{i}\alpha_1)$, $\xi_1=\xi_{1R}+{\rm i}\xi_{1I}$, the suffixes $R$ and $I$ denote the real and imaginary parts, respectively. Obviously, the amplitude of the bright soliton in the $\Phi^{(k)}$ component is $\frac{|c^{(k)}_1|}{2} {\rm e}^{-\eta_1}$ while the amplitude of the bright soliton in the $u$ component is $2p^2_{1R}$. For the dark soliton in the $\Phi^{(3)}$ component, $|\Phi^{(3)}|$ approaches $|\rho_1|$ as $x,y \rightarrow \pm \infty$. In addition, the intensity of the dark soliton is $|\rho_1|\cos\phi_1$.
As the parameters $c^{(k)}_1$ appear explicitly in the amplitude of the bright parts of the mixed one-soliton, we can tune the intensity of bright parts without altering the depth of the dark part.
The mixed one-soliton at time $t=0$ is displayed in Fig.\ref{mix-fig1} with the nonlinearities $(\sigma_1,\sigma_2,\sigma_3)=(1,-1,1)$.
The parameters used are $p_1=1+\frac{1}{2}{\rm i}$, $\alpha_1=\rho_1=1$, $\xi_{10}=y=0$, $c^{(2)}_1=1+{\rm i}$ and (a) $c^{(1)}_1=2$; (b) $c^{(1)}_1=2+3{\rm i}$.
It can be seen that when the parameters $c^{(k)}_1$ take different values, the intensities of the bright solitons in the $\Phi^{(1)}$ and $\Phi^{(2)}$ components change, while the intensity of the bright soliton in the $u$ component and the depth of the dark soliton remain unaltered.


\subsubsection{Two-soliton solution}
To obtain two-soliton solution, we take $N=2$ in the formula (\ref{j44}).
In this case, the tau functions can be rewritten as
\begin{align}
& f=1+E_{11^*}\textmd{e}^{\xi_1+\xi^*_1}+E_{12^*}\textmd{e}^{\xi_1+\xi^*_2}+E_{21^*}\textmd{e}^{\xi_2+\xi^*_1}+E_{22^*}\textmd{e}^{\xi_2+\xi^*_2}+E_{121^*2^*}\textmd{e}^{\xi_1+\xi_2+\xi^*_1+\xi^*_2},\label{jj49}\\
& g^{(k)}=c^{(k)}_1\textmd{e}^{\xi_1}+c^{(k)}_2\textmd{e}^{\xi_2}+G^{(k)}_{121^*}\textmd{e}^{\xi_1+\xi_2+\xi^*_1}+G^{(k)}_{122^*}\textmd{e}^{\xi_1+\xi_2+\xi^*_2},\ \ \ \ \ k=1,2,\label{jj50}\\
& h^{(1)}=1+F^{(1)}_{11^*}\textmd{e}^{\xi_1+\xi^*_1}+F^{(1)}_{1,2^*}\textmd{e}^{\xi_1+\xi^*_2}+F^{(1)}_{21^*}\textmd{e}^{\xi_2+\xi^*_1}+F^{(1)}_{22^*}\textmd{e}^{\xi_2+\xi^*_2}+F^{(1)}_{121^*2^*}\textmd{e}^{\xi_1+\xi_2+\xi^*_1+\xi^*_2},\label{jj51}
\end{align}
where
\begin{align*}
& E_{ij^*}=\Big(\sum^2_{k=1}\sigma_kc_i^{(k)}c_j^{(k)*}\Big)\Big[8(p_i+p_j^*)(p_i^3+{p_j^*}^3)+\frac{\sigma_3\rho^2_1(p_i+p_j^*)^2}{(p_i-\textmd{i}\alpha_1)(p_j^*+\textmd{i}\alpha_1)}\Big]^{-1},\\
& E_{121^*2^*}=|p_1-p_2|^2\Big[\frac{E_{11^*}E_{22^*}}{(p_1+p_2^*)(p_2+p_1^*)}-\frac{E_{12^*}E_{21^*}}{(p_1+p_1^*)(p_2+p_2^*)}\Big],\\
& F^{(1)}_{ij^*}=-\frac{p_i-\textmd{i}\alpha_1}{p^*_j+\textmd{i}\alpha_1} E_{ij^*},\\
& F^{(1)}_{121^*2^*}=\frac{(p_1-\textmd{i}\alpha_1)(p_2-\textmd{i}\alpha_1)}{(p^*_1+\textmd{i}\alpha_1)(p^*_2+\textmd{i}\alpha_1)}E_{121^*2^*},\\
& G^{(k)}_{12i^*}=(p_1-p_2)\Big(\frac{c^{(k)}_1E_{2i^*}}{p_1+p_i^*}-\frac{c^{(k)}_2E_{1i^*}}{p_2+p_i^*}\Big),
\end{align*}
with
$ \xi_i=p_ix -\textmd{i}p^2_iy-8p^3_i t  + \xi_{i0}$ for $i=1,2$.

To get nonsingular solution, the denominator $f$ must be nonzero. For this purpose, we rewrite $f$ as
\begin{align}
& f=2\textmd{e}^{\xi_{1R}+\xi_{2R}}[\textmd{e}^{\zeta_{R}} \cos(\xi_{1I}-\xi_{2I}+\zeta_{I})+\textmd{e}^{\eta_1+\eta_2} \cosh(\xi_{1R}-\xi_{2R}+\eta_1-\eta_2)+\textmd{e}^{\eta_3} \cosh(\xi_{1R}-\xi_{2R}+\eta_3)],
\end{align}
where
\begin{align*}
& \textmd{e}^{2\eta_1}=E_{11^*},\ \ \ \ \textmd{e}^{2\eta_2}=E_{22^*},\ \ \ \ \textmd{e}^{2\eta_3}=E_{121^*2^*},\ \ \ \ \textmd{e}^{\zeta_{R}+\textmd{i}\zeta_{I}}=E_{12^*}.
\end{align*}
Combine the condition for the existence of one-soliton solution, it is easy to known that $E_{ii^*}>0,i=1,2$ is a necessary condition
and ${\rm e}^{\eta_1+\eta_2}+{\rm e}^{\eta_3}>{\rm e}^{\zeta_{R}}$ is a sufficient condition to guarantee a
nonsingular two-soliton solution.
Based on the above tau functions, the asymptotic analysis of the two-soliton solution can be performed as in Refs. \cite{lak1, lak3}, this will help us to elucidate the understanding of the collision of two solitons. Particularly, we study the interaction of two solitons in the $x$-$y$ plane. The similar approach can also be applied to study the collision in other planes.

The asymptotic forms of two solitons $s_1$ and $s_2$ before and after collision can be deduced from the transformations (\ref{jj7}) and the tau functions (\ref{jj49})-(\ref{jj51}).

(a) Before collision ($x,y\rightarrow -\infty$)

Soliton $s_1$
\begin{align*}
& \Phi_{1-}^{(k)} \simeq A^{(k)}_{1-}\textmd{e}^{\textmd{i}\xi_{1I}}{\rm sech}(\xi_{1R}+\eta_1),\ \ \ \ \ k=1,2,\\
& \Phi_{1-}^{(3)} \simeq \frac{\rho_1}{2} \textmd{e}^{\textmd{i}\theta_{1}} [1+\textmd{e}^{2\textmd{i}\phi_1}+(\textmd{e}^{2\textmd{i}\phi_1}-1)\tanh(\xi_{1R}+\eta_1)],\\
& u_{1-} \simeq 2p^2_{1R}{\rm sech}^2(\xi_{1R}+\eta_1).
\end{align*}

Soliton $s_2$
\begin{align*}
& \Phi_{2-}^{(k)} \simeq A^{(k)}_{2-}\textmd{e}^{\textmd{i}\xi_{2I}}{\rm sech}(\xi_{2R}+\eta_3-\eta_1),\ \ \ \ \ k=1,2,\\
& \Phi_{2-}^{(3)} \simeq \frac{\rho_1}{2} \textmd{e}^{\textmd{i}(\theta_1+2\phi_1)}[1+\textmd{e}^{2\textmd{i}\phi_2}+(\textmd{e}^{2\textmd{i}\phi_2}-1)\tanh(\xi_{2R}+\eta_3-\eta_1)],\\
& u_{2-} \simeq 2p^2_{2R}{\rm sech}^2(\xi_{2R}+\eta_3-\eta_1).
\end{align*}

(b) After collision ($x,y\rightarrow +\infty$)

Soliton $s_1$
\begin{align*}
& \Phi_{1+}^{(k)} \simeq A^{(k)}_{1+}\textmd{e}^{\textmd{i}\xi_{1I}}{\rm sech}(\xi_{1R}+\eta_3-\eta_2),\ \ \ \ \ k=1,2,\\
& \Phi_{1+}^{(3)} \simeq \frac{\rho_1}{2} \textmd{e}^{\textmd{i}(\theta_1+2\phi_2)}[1+\textmd{e}^{2\textmd{i}\phi_1}+(\textmd{e}^{2\textmd{i}\phi_1}-1)\tanh(\xi_{1R}+\eta_3-\eta_2)],\\
& u_{1+} \simeq 2p^2_{1R}{\rm sech}^2(\xi_{1R}+\eta_3-\eta_2).
\end{align*}

Soliton $s_2$
\begin{align*}
& \Phi_{2+}^{(k)} \simeq A^{(k)}_{2+}\textmd{e}^{\textmd{i}\xi_{2I}}{\rm sech}(\xi_{2R}+\eta_2),\ \ \ \ \ k=1,2,\\
& \Phi_{2+}^{(3)} \simeq \frac{\rho_1}{2} \textmd{e}^{\textmd{i}\theta_1}[1+\textmd{e}^{2\textmd{i}\phi_2}+(\textmd{e}^{2\textmd{i}\phi_2}-1)\tanh(\xi_{2R}+\eta_2)],\\
& u_{2+} \simeq 2p^2_{2R}{\rm sech}^2(\xi_{2R}+\eta_2).
\end{align*}

In the above expressions, $\textmd{e}^{2\textmd{i}\phi_j}=-(p_j-\textmd{i}\alpha_1)/(p^*_j+\textmd{i}\alpha_1)$ for $j=1,2$; $( A^{(1)}_{1-}, A^{(1)}_{2-})$ are the amplitudes of the bright parts of the mixed two solitons $s_1$ and $s_2$ before interaction; $( A^{(1)}_{1+}, A^{(1)}_{2+})$ are the corresponding amplitudes after interaction. Where the superscript (subscript) of $A$ represents the component (soliton) number and $-\ (+)$ denotes the soliton before (after) collision, and the various amplitudes are given by
\begin{align*}
& A^{(k)}_{1-}=\frac{c^{(k)}_1}{2\sqrt{E_{11^*}}},\ \ \ \ \  A^{(k)}_{2-}=\frac{G^{(k)}_{121^*}}{2\sqrt{E_{11^*}E_{121^*2^*}}}, \ \ \ \ \ A^{(k)}_{1+}=\frac{G^{(k)}_{122^*}}{2\sqrt{E_{22^*}E_{121^*2^*}}},\ \ \ \ \ A^{(k)}_{2+}=\frac{c^{(k)}_2}{2\sqrt{E_{22^*}}}.
\end{align*}
What's more, there is a relation between the amplitudes before and after collision \cite{lak3}
\begin{align}\label{jj53}
& A^{(k)}_{i+}=T_i^{k}A^{(k)}_{i-},\ \ \ \ \ i,k=1,2,
\end{align}
where the transition amplitudes $T_i^{k}$ are given by
\begin{align*}
& T_1^{k}=\Big(\frac{p_1-p_2}{p^*_1-p^*_2}\Big)\Big(\frac{p^*_1+p_2}{p_1+p^*_2}\Big)^{1/2}\Big[\frac{(c^{(k)}_2/c^{(k)}_1)r_1-1}{\sqrt{1-r_1r_2}}\Big],\ \ \
 T_2^{k}=\Big(\frac{p_1-p_2}{p^*_1-p^*_2}\Big)\Big(\frac{p^*_1+p_2}{p_1+p^*_2}\Big)^{1/2}\Big[\frac{\sqrt{1-r_1r_2}}{(c^{(k)}_1/c^{(k)}_2)r_2-1}\Big],\ \ \ k=1,2,
\end{align*}
with
\begin{align*}
& r_1=\frac{p_1+p^*_2}{p_2+p^*_2}\frac{E_{12^*}}{E_{22^*}},\ \ \ \ \ r_2=\frac{p_1^*+p_2}{p_1+p^*_1}\frac{E_{21^*}}{E_{11^*}}.
\end{align*}
The relation (\ref{jj53}) means that the intensities of the bright parts of the mixed two solitons before and after collision alter in general. Only under the condition $\frac{|c^{(1)}_1|}{|c^{(1)}_2|}=\frac{|c^{(2)}_1|}{|c^{(2)}_2|}$, the transition amplitudes $T_i^{k}$ become unimodular. It indicates that in general, the bright parts of the mixed two solitons exhibit energy-exchanging inelastic collision characterized with an intensity redistribution (energy sharing) among the bright parts of the mixed solitons in the $\Phi^{(1)}$ and $\Phi^{(2)}$ components.
As the amplitudes of the dark parts of the mixed two solitons in the $\Phi^{(3)}$ component are same and equal to $\rho_1$, the intensities of the dark parts remain unaltered after collision. Thus the dark parts of the mixed two solitons undergo mere elastic collision.
What's more, both the bright and dark parts of the mixed two solitons admit a same magnitude position shift but with opposite signs.
The position shift of soliton $s_1$ (and $s_2$) is $\Lambda_1=\eta_3-\eta_1-\eta_2$ (and $\Lambda_2=-\Lambda_1$).
In addition, the similar analysis of the $u$ component shows that standard elastic collision takes place between the two bright solitons in the $u$ component.
The collisions of two solitons of the three-component Mel'nikov system (\ref{j5})-(\ref{j8}) with the nonlinearities $(\sigma_1,\sigma_2,\sigma_3)=(1,-1,1)$ are depicted in Figs.\ref{mix-fig2} and \ref{mix-fig3} at the time $t=0$. The parameters used in Fig.\ref{mix-fig2} are $p_1=\frac{2}{5}+\frac{1}{5}{\rm i}, p_2=\frac{4}{5}-\frac{2}{5}{\rm i}, c^{(1)}_1=1+\frac{1}{2}{\rm i},c^{(1)}_2=\frac{5}{2}, c^{(2)}_1=\frac{1}{2}{\rm i},c^{(2)}_2=\frac{1}{4}-\frac{1}{4}{\rm i},\rho_1=1, \alpha_1=\frac{1}{2}$
and $ \xi_{10}=\xi_{20}=0$, which correspond to inelastic collisions of the bright solitons in the $\Phi^{(1)}$ and $\Phi^{(2)}$ components.
An example of elastic collision of the bright solitons in the $\Phi^{(1)}$ and $\Phi^{(2)}$ components is displayed in Fig.\ref{mix-fig3}
with the parameters $c^{(1)}_1,c^{(1)}_2, c^{(2)}_1,c^{(2)}_2$ used in Fig.\ref{mix-fig2} are replaced by $ c^{(1)}_1=1,c^{(1)}_2=2, c^{(2)}_1=\frac{1}{3},c^{(2)}_2=\frac{2}{3}$.
In Figs.\ref{mix-fig2} and \ref{mix-fig3}, (a) and (b) represent the collisions of two bright solitons in the $\Phi^{(1)}$ and $\Phi^{(2)}$ components, respectively; the collision of two dark solitons in the $\Phi^{(3)}$ component is displayed in (c); and (d) represents the collision of two bright solitons in the $u$ component. The difference between Figs.\ref{mix-fig2} and \ref{mix-fig3} is inelastic and elastic collisions, which only appears in (a) and (b).
It is obvious that in the (a) and (b) of Figs.\ref{mix-fig2}, the intensity of a soliton is suppressed while the intensity of the other soliton is enhanced after collision. The physical mechanism behind this interesting collision is attributed to an intensity redistribution among the components accompanied by finite amplitude-dependent position shift.



\subsection{1-b-2-d mixed soliton solution}

In this case, assuming the $\Phi^{(1)}$ component is of bright type while the $\Phi^{(2)}$ and $\Phi^{(3)}$ components are of dark type, we introduce the dependent variable transformations
\begin{align}
& \Phi^{(1)}= \frac{g^{(1)}}{f},\ \ \ \ \Phi^{(2)}=\rho_1 {\rm e}^{\textmd{i}\theta_1} \frac{h^{(1)}}{f},\ \ \ \ \Phi^{(3)}=\rho_2 {\rm e}^{\textmd{i}\theta_2} \frac{h^{(2)}}{f},\ \ \ \ u=2 (\log f)_{xx},
\end{align}
which convert the three-component Mel'nikov system (\ref{j5})-(\ref{j8}) into the bilinear forms
\begin{align}
& (D^2_x-{\rm i} D_y)g^{(1)} \cdot f=0,\label{j61}\\
& (D^2_x+2 \textmd{i} \alpha_k D_x-{\rm i} D_y)h^{(k)} \cdot f=0,\ \ \ \ \ \ k=1,2,\label{j62}\\
& (D^4_x+D_xD_t-3D^2_y)f\cdot f=-\sigma_1 g^{(1)}{g^{(1)}}^\ast+\sum^2_{k=1}\sigma_{k+1} \rho^2_k (f^2-h^{(k)}{h^{(k)}}^\ast),\label{j63}
\end{align}
where $f$ is a real function; $g^{(1)}, h^{(1)}$ and $h^{(2)}$ are complex functions; $\theta_k=\alpha_kx+\alpha^2_ky+\beta_k(t)$, $\alpha_k$ and $\rho_k , (k=1,2)$ are real constants, $\beta_k(t) , (k=1,2)$ are real functions.

To construct the 1-b-2-d mixed $N$-soliton solution, we start with a two-component KP hierarchy with two copies of shifted singular points ($c_1$ and $c_2$)
\begin{align}
& (D^2_{x_1}-D_{x_2})\tau_{1}(k_1,k_2) \cdot \tau_{0}(k_1,k_2)=0,\label{j68}\\
& (D^2_{x_1}-D_{x_2}+2c_1D_{x_1})\tau_{0}(k_1+1,k_2) \cdot \tau_{0}(k_1,k_2)=0,\\
& (D^2_{x_1}-D_{x_2}+2c_2D_{x_1})\tau_{0}(k_1,k_2+1) \cdot \tau_{0}(k_1,k_2)=0,\\
& (D^4_{x_1}-4D_{x_1}D_{x_3}+3D^2_{x_2})\tau_{0}(k_1,k_2) \cdot \tau_{0}(k_1,k_2)=0,\\
& D_{x_1}D_{y^{(1)}_1}\tau_{0}(k_1,k_2) \cdot \tau_{0}(k_1,k_2)=-2\tau_{1}(k_1,k_2) \tau_{-1}(k_1,k_2),\\
& (D_{x_1}D_{x^{(1)}_{-1}}-2)\tau_{0}(k_1,k_2) \cdot \tau_{0}(k_1,k_2)=-2\tau_{0}(k_1+1,k_2) \tau_{0}(k_1-1,k_2),\\
& (D_{x_1}D_{x^{(2)}_{-1}}-2)\tau_{0}(k_1,k_2) \cdot \tau_{0}(k_1,k_2)=-2\tau_{0}(k_1,k_2+1) \tau_{0}(k_1,k_2-1).\label{j74}
\end{align}
Based on the Sato theory for KP hierarchy \cite{jimbo1}, the bilinear equations (\ref{j68})-(\ref{j74}) have the Gram determinant tau function solution
\begin{align}
& \tau_{0}(k_1,k_2)=\left| \begin{array}{ccccc}
A & I  \\
-I & B
\end{array} \right|,\\
& \tau_{1}(k_1,k_2)=\left| \begin{array}{ccccc}
A & I  & \Omega^{\textmd{T}}\\
-I & B & \mathbf{0}^{\textmd{T}}\\
\mathbf{0} & -\bar{\Psi} & 0
\end{array} \right|,\ \ \ \ \ \
 \tau_{-1}(k_1,k_2)=\left| \begin{array}{ccccc}
A & I  & \mathbf{0}^{\textmd{T}}\\
-I & B & \Psi^{\textmd{T}}\\
-\bar{\Omega} & \mathbf{0} & 0
\end{array} \right|,
\end{align}
where $\Omega,\Psi,\bar{\Omega},\bar{\Psi}$  are the $N$-component row vectors defined previously; $A$ and $B$ are $N \times N$ matrices whose entries are
\begin{align*}
& a_{ij}(k_1,k_2)=\frac{1}{p_i+\bar{p}_j} \Big(-\frac{p_i-c_1}{\bar{p}_j+c_1}\Big)^{k_1} \Big(-\frac{p_i-c_2}{\bar{p}_j+c_2}\Big)^{k_2} \textmd{e}^{\xi_i+\bar{\xi}_j},\ \ \ \ \
 b_{ij}=\frac{1}{q_i+\bar{q}_j}\textmd{e}^{\eta_i+\bar{\eta}_j},
\end{align*}
with
\begin{align*}
& \xi_i=\frac{1}{p_i-c_1}x^{(1)}_{-1}+\frac{1}{p_i-c_2}x^{(2)}_{-1}+p_ix_1+p^2_ix_2+p^3_ix_3+\xi_{i0},\\
& \bar{\xi}_j=\frac{1}{\bar{p}_j+c_1}x^{(1)}_{-1}+\frac{1}{\bar{p}_j+c_2}x^{(2)}_{-1}+\bar{p}_jx_1-\bar{p}^2_jx_2+\bar{p}^3_jx_3+\bar{\xi}_{j0},\\
& \eta_i=q_iy^{(1)}_1+\eta_{i0},\ \ \ \ \ \ \ \ \ \ \bar{\eta}_j=\bar{q}_jy^{(1)}_1+\bar{\eta}_{j0},
\end{align*}
in which $p_i, \bar{p}_j, q_i, \bar{q}_j, \xi_{i0}, \bar{\xi}_{j0}, \eta_{i0}, \bar{\eta}_{j0}, c_1$ and $c_2$ are complex constants.

We also first consider complex conjugate reduction by assuming $x_1$, $x_3$, $x^{(1)}_{-1}$, $x^{(2)}_{-1}$, $y^{(1)}_1$ are real; $x_2$, $c_1$ and $c_2$ are pure imaginary and defining $p^*_j=\bar{p}_j$, $q^*_j=\bar{q}_j$, $\xi^*_{j0}=\bar{\xi}_{j0}$ and $\eta^*_{j0}=\bar{\eta}_{j0}$.
It is easy to check that
\begin{align*}
a_{ij}(k_1,k_2)=a^*_{ji}(-k_1,-k_2),\ \ \ \ \ \ \ \ b_{ij}=b^*_{ji}.
\end{align*}
Furthermore, by letting
\begin{align*}
f=\tau_{0}(0,0),\ \ \  g^{(1)}=\tau_{1}(0,0),\ \ \  h^{(1)}=\tau_{0}(1,0),\ \ \  h^{(2)}=\tau_{0}(0,1),
\end{align*}
thus, $f$ is real and
\begin{align*}
g^{(1)*}=-\tau_{-1}(0,0),\ \ \  h^{(1)*}=\tau_{0}(-1,0),\ \ \  h^{(2)*}=\tau_{0}(0,-1).
\end{align*}
Then the bilinear equations (\ref{j68})-(\ref{j74}) become to
\begin{align}
& (D^2_{x_1}-D_{x_2})g^{(1)} \cdot f=0,\label{j75}\\
& (D^2_{x_1}-D_{x_2}+2c_kD_{x_1})h^{(k)} \cdot f=0,\ \ \ \ k=1,2,\label{j76}\\
& (D^4_{x_1}-4D_{x_1}D_{x_3}+3D^2_{x_2})f\cdot f=0,\label{j77}\\
& D_{x_1}D_{y^{(1)}_1}f \cdot f=2g^{(1)}g^{(1)*},\label{j78}\\
& (D_{x_1}D_{x^{(k)}_{-1}}-2)f \cdot f=-2h^{(k)} h^{(k)*},\ \ \ \ \ k=1,2.\label{j79}
\end{align}
Also by the transformations (\ref{j42}), equations (\ref{j75})-(\ref{j76}) are recast into equations (\ref{j61})-(\ref{j62}) along with $c_1={\rm i}\alpha_1$ and $c_2={\rm i}\alpha_2$.
Thus, the remaining task is to derive equation (\ref{j63}) from equations (\ref{j77})-(\ref{j79}).

In a similar way, under the reduction conditions
\begin{align}
& 8{p_i^*}^3=\sigma_1q_i-\frac{\sigma_2\rho^2_1}{p_i^*+c_1}-\frac{\sigma_3\rho^2_2}{p_i^*+c_2},\ \ \ \ \ \ \  8p^3_j=\sigma_1q^*_j-\frac{\sigma_2\rho^2_1}{p_j-c_1}-\frac{\sigma_3\rho^2_2}{p_j-c_2},
\end{align}
i.e.,
\begin{align}
& \frac{1}{q_i+q^*_j}=\frac{\sigma_1}{8({p_i^*}^3+p_j^3)+\frac{\sigma_2\rho^2_1(p_i^*+p_j)}{(p_i^*+c_1)(p_j-c_1)}+\frac{\sigma_3\rho^2_2(p_i^*+p_j)}{(p_i^*+c_2)(p_j-c_2)}},
\end{align}
the following relation holds
\begin{align}
& 8f_{x_3}=\sigma_1f_{y^{(1)}_1}-\sigma_2\rho^2_1f_{x^{(1)}_{-1}}-\sigma_3\rho^2_2f_{x^{(2)}_{-1}},\label{j82}
\end{align}
which also gives
\begin{align}
& 8f_{x_1x_3}=\sigma_1f_{x_1y^{(1)}_1}-\sigma_2\rho^2_1f_{x_1x^{(1)}_{-1}}-\sigma_3\rho^2_2f_{x_1x^{(2)}_{-1}}.\label{j83}
\end{align}

Similarly, equations (\ref{j78}) and (\ref{j79}) can be expanded as
\begin{align}
& f_{x_1y^{(1)}_1}f-f_{x_1}f_{y^{(1)}_{1}}=g^{(1)}g^{(1)*},\label{j84}
\end{align}
and
\begin{align}
& f_{x_1x^{(1)}_{-1}}f-f_{x_1}f_{x^{(1)}_{-1}}-f^2=-h^{(1)} h^{(1)*},\ \ \ \ \ \ \ \ \ f_{x_1x^{(2)}_{-1}}f-f_{x_1}f_{x^{(2)}_{-1}}-f^2=-h^{(2)} h^{(2)*},\label{j85}
\end{align}
respectively. By referring to the relations (\ref{j82}) and (\ref{j83}), from (\ref{j84}) and (\ref{j85}), we can arrive at
\begin{align}
& -4D_{x_1}D_{x_3}f\cdot f=-\sigma_1g^{(1)}g^{(1)*}+\sigma_2\rho^2_1(f^2-h^{(1)}h^{(1)*})+\sigma_3\rho^2_2(f^2-h^{(2)}h^{(2)*}).\label{j86}
\end{align}
By using the transformations (\ref{j42}), from equations (\ref{j77}) and (\ref{j86}), equation (\ref{j63}) is immediately obtained.

In conclude, we have obtained the 1-b-2-d mixed $N$-soliton solution to the three-component Mel'nikov system (\ref{j5})-(\ref{j8})
\begin{align}\label{j87}
& f=\left| \begin{array}{ccccc}
A & I  \\
-I & B
\end{array} \right|,\ \ \ \ \ \ \
 g^{(1)}=\left| \begin{array}{ccccc}
A & I  & \Omega^{\textmd{T}}\\
-I & B & \mathbf{0}^{\textmd{T}}\\
\mathbf{0} & C_1 & 0
\end{array} \right|,\ \ \ \ \ \ \
h^{(k)}=\left| \begin{array}{ccccc}
A^{(k)} & I  \\
-I & B
\end{array} \right|,
\end{align}
where the elements in $A,A^{(k)}$ and $B$ are given by
\begin{align}
& a_{ij}=\frac{1}{p_i+p^*_j} \textmd{e}^{\xi_i+\xi^*_j},\ \ \ \  \ \  a^{(k)}_{ij}=\frac{1}{p_i+p^*_j} \Big(-\frac{p_i-\textmd{i}\alpha_k}{p^*_j+\textmd{i}\alpha_k}\Big) \textmd{e}^{\xi_i+\xi^*_j},\\ &b_{ij}=\sigma_1c_i^{(1)*}c_j^{(1)}\Big[8({p_i^*}^3+p_j^3)+\sum^2_{l=1}\frac{\sigma_{l+1}\rho^2_l(p_i^*+p_j)}{(p_i^*+\textmd{i}\alpha_l)(p_j-\textmd{i}\alpha_l)}\Big]^{-1},\label{j88}
\end{align}
meanwhile, $\Omega$ and $C_1$ are
\begin{align}
& \Omega=(\textmd{e}^{\xi_1},\textmd{e}^{\xi_2},\cdots,\textmd{e}^{\xi_N}),\ \ \ \ \ \ C_1=-(c^{(1)}_1,c^{(1)}_2,\cdots,c^{(1)}_N),
\end{align}
with $ \xi_i=p_ix -\textmd{i}p^2_iy-8p^3_i t  + \xi_{i0}$; $p_i$, $\xi_{i0}$ and $c^{(1)}_i$, $(k=1,2; i=1,2,\cdots,N)$ are complex constants.

\subsubsection{One-soliton solution}
To get one-soliton solution, we take $N=1$ in the formula (\ref{j87}).
In this case, the tau functions can be rewritten as
\begin{align}
& f=1+E_{11^*}\textmd{e}^{\xi_1+\xi^*_1},\\
& g^{(1)}=c_1^{(1)}\textmd{e}^{\xi_1},\\
& h^{(k)}=1+F_{11^*}\textmd{e}^{\xi_1+\xi^*_1},\ \ \ \ \ k=1,2,
\end{align}
where
\begin{align*}
& E_{11^*}=\sigma_1c_1^{(1)}c_1^{(1)*}\Big[8(p_1+p_1^*)(p_1^3+{p_1^*}^3)+\sum^2_{l=1}\frac{\sigma_{l+1}\rho^2_l(p_1+p_1^*)^2}{(p_1-\textmd{i}\alpha_l)(p_1^*+\textmd{i}\alpha_l)}\Big]^{-1},\\
& F_{11^*}=-\frac{p_1-\textmd{i}\alpha_k}{p^*_1+\textmd{i}\alpha_k}E_{11^*},
\end{align*}
with $\xi_1=p_1x -\textmd{i}p^2_1y-8p^3_1 t  + \xi_{10}$.
It is worth noting that this solution is nonsingular only when $E_{11^*}>0$.

The 1-b-2-d mixed one-soliton solution can be expressed in the following form
\begin{align}
& \Phi^{(1)}=\frac{c^{(1)}_1}{2}  \textmd{e}^{\textmd{i}\xi_{1I}-\eta_1} {\rm sech}[\xi_{1R}+\eta_1],\\
& \Phi^{(k+1)}=\frac{\rho_k}{2} \textmd{e}^{\textmd{i}\theta_{k}} \{1+\textmd{e}^{2\textmd{i}\phi_k}+(\textmd{e}^{2\textmd{i}\phi_k}-1)\tanh[\xi_{1R}+\eta_1]\},\ \ \ \ \ k=1,2,\\
& u=2p^2_{1R}{\rm sech}^2[\xi_{1R}+\eta_1],
\end{align}
where $\textmd{e}^{2 \eta_1}=E_{11^*}$, $\textmd{e}^{2\textmd{i}\phi_k}=-(p_1-\textmd{i}\alpha_k)/(p^*_1+\textmd{i}\alpha_k)$, $\xi_1=\xi_{1R}+{\rm i}\xi_{1I}$.
The amplitude of the bright part of the mixed single soliton in the $\Phi^{(1)}$ component is $\frac{|c^{(1)}_1|}{2} {\rm e}^{-\eta_1}$ while the amplitude of the bright soliton in the $u$ component is $2p^2_{1R}$.
For the dark parts of the mixed single soliton in the $\Phi^{(2)}$ and $\Phi^{(3)}$ components, $|\Phi^{(k+1)}|,  (k=1,2)$ approaches $|\rho_k|$ as $x,y \rightarrow \pm \infty$.
What's more, the intensity of the dark soliton in the $\Phi^{(k+1)}$ component is $|\rho_k|\cos\phi_k$ for $k=1,2$.
In addition, according to the values of $\alpha_1$ and $\alpha_2$, there exist two different cases:
(i) $\alpha_1=\alpha_2$ and (ii) $\alpha_1\neq\alpha_2$.
In the former case, we have $\phi_1=\phi_2$, which implies that the dark solitons in the $\Phi^{(2)}$ and $\Phi^{(3)}$ components are proportional to each other.
Therefore, this case is considered as degenerate.
In the latter case, the condition $\phi_1\neq\phi_2$ means that the dark solitons in the $\Phi^{(2)}$ and $\Phi^{(3)}$ components
have different degrees of darkness at the center.
In this case, $\Phi^{(2)}$ and $\Phi^{(3)}$ are not proportional to each other, thus it is viewed as non-degenerate case.
Both the degenerate and non-degenerate cases are illustrated in Fig.\ref{mix-fig4} with the parametric choice $p_1=1+\frac{1}{2}{\rm i}$, $c^{(1)}_1=\rho_1=1$, $\xi_{10}=y=0$ and (a) $\alpha_1=\alpha_2=\frac{1}{2}$, $\rho_2=2$; (b) $\alpha_1=2, \alpha_2=-\frac{1}{3}$, $\rho_2=1$ at time $t=0$ with the nonlinearities $(\sigma_1,\sigma_2,\sigma_3)=(1,-1,1)$.


\subsubsection{Two-soliton solution}
To get two-soliton solution, we take $N=2$ in the formula (\ref{j87}).
In this case, the corresponding tau functions can be rewritten as
\begin{align}
& f=1+E_{11^*}\textmd{e}^{\xi_1+\xi^*_1}+E_{12^*}\textmd{e}^{\xi_1+\xi^*_2}+E_{21^*}\textmd{e}^{\xi_2+\xi^*_1}+E_{22^*}\textmd{e}^{\xi_2+\xi^*_2}+E_{121^*2^*}\textmd{e}^{\xi_1+\xi_2+\xi^*_1+\xi^*_2},\\
& g^{(1)}=c^{(1)}_1\textmd{e}^{\xi_1}+c^{(1)}_2\textmd{e}^{\xi_2}+G^{(1)}_{121^*}\textmd{e}^{\xi_1+\xi_2+\xi^*_1}+G^{(1)}_{122^*}\textmd{e}^{\xi_1+\xi_2+\xi^*_2},\\
& h^{(k)}=1+F^{(k)}_{11^*}\textmd{e}^{\xi_1+\xi^*_1}+F^{(k)}_{12^*}\textmd{e}^{\xi_1+\xi^*_2}+F^{(k)}_{21^*}\textmd{e}^{\xi_2+\xi^*_1}+F^{(k)}_{22^*}\textmd{e}^{\xi_2+\xi^*_2} +F^{(k)}_{121^*2^*}\textmd{e}^{\xi_1+\xi_2+\xi^*_1+\xi^*_2},\ \ \ \ \ k=1,2,
\end{align}
where
\begin{align*}
& E_{ij^*}=\sigma_1c_i^{(1)}c_j^{(1)*}\Big[8(p_i+p_j^*)(p_i^3+{p_j^*}^3)+\sum^2_{l=1}\frac{\sigma_{l+1}\rho^2_l(p_i+p_j^*)^2}{(p_i-\textmd{i}\alpha_l)(p_j^*+\textmd{i}\alpha_l)}\Big]^{-1},\\
& F^{(k)}_{ij^*}=-\frac{p_i-\textmd{i}\alpha_k}{p^*_j+\textmd{i}\alpha_k} E_{ij^*},\\
& E_{121^*2^*}=|p_1-p_2|^2\Big[\frac{E_{11^*}E_{22^*}}{(p_1+p_2^*)(p_2+p_1^*)}-\frac{E_{12^*}E_{21^*}}{(p_1+p_1^*)(p_2+p_2^*)}\Big],\\
& F^{(k)}_{121^*2^*}=\frac{(p_1-\textmd{i}\alpha_k)(p_2-\textmd{i}\alpha_k)}{(p^*_1+\textmd{i}\alpha_k)(p^*_2+\textmd{i}\alpha_k)}E_{121^*2^*},\\
& G^{(1)}_{12i^*}=(p_1-p_2)\Big(\frac{c^{(1)}_1E_{2i^*}}{p_1+p_i^*}-\frac{c^{(1)}_2E_{1i^*}}{p_2+p_i^*}\Big),
\end{align*}
with
$ \xi_i=p_ix -\textmd{i}p^2_iy-8p^3_i t  + \xi_{i0}$ for $i=1,2$.

Similarly, nonsingular solution requires the denominator $f$ must be nonzero.
Also, we rewrite $f$ as
\begin{align}
& f=2\textmd{e}^{\xi_{1R}+\xi_{2R}}[\textmd{e}^{\zeta_{R}} \cos(\xi_{1I}-\xi_{2I}+\zeta_{I})+\textmd{e}^{\eta_1+\eta_2} \cosh(\xi_{1R}-\xi_{2R}+\eta_1-\eta_2)+\textmd{e}^{\eta_3} \cosh(\xi_{1R}-\xi_{2R}+\eta_3)],
\end{align}
where $\textmd{e}^{2\eta_1}=E_{11^*}, \textmd{e}^{2\eta_2}=E_{22^*},\textmd{e}^{2\eta_3}=E_{121^*2^*}, \textmd{e}^{\zeta_{R}+\textmd{i}\zeta_{I}}=E_{12^*}$. As in the above subsection, it is easy to know that $E_{ii^*}>0, (i=1,2)$ is a necessary condition and ${\rm e}^{\eta_1+\eta_2}+{\rm e}^{\eta_3}>{\rm e}^{\zeta_{R}}$ is a sufficient condition to guarantee a nonsingular two-soliton solution.

Following the method in the previous subsection, the similar asymptotic analysis of the collision between two solitons also can be performed, whose details are omitted here. The amplitudes of the bright parts of the mixed two solitons before and after collision are given by
\begin{align}\label{jj92}
& (A^{(1)}_{1-},A^{(1)}_{2-},A^{(1)}_{1+},A^{(1)}_{2+})=\Big(\frac{c^{(1)}_1}{2\sqrt{E_{11^*}}}, \frac{G^{(1)}_{121^*}}{2\sqrt{E_{11^*}E_{121^*2^*}}}, \frac{G^{(1)}_{122^*}}{2\sqrt{E_{22^*}E_{121^*2^*}}},\frac{c^{(1)}_2}{2\sqrt{E_{22^*}}}\Big).
\end{align}
After substituting the expressions of various quantities, a tedious calculation shows that $|A^{(1)}_{j-}|=|A^{(1)}_{j+}|$ for $j=1,2$, which indicates that the intensities of the bright parts of the colliding mixed two solitons are same before and after collision. Similarly, the amplitudes of the dark parts of the mixed two solitons in the $\Phi^{(k+1)}$ component before and after interaction are same and equal to $\rho_k$. In addition, it also can be seen that the bright solitons in the $u$ component also undergo elastic collision.
Hence, different from the 2-b-1-d case, the two solitons in all the components (short-wave and long-wave components) for the 1-b-2-d case always undertake standard elastic collision without energy exchange. It is worth noting that this phenomenon is consistent with the mixed soliton solution of the three-component YO system \cite{lak3,chen2}. The mixed-type two solitons are displayed in Fig.\ref{mix-fig5} with the parametric choice $p_1=\frac{2}{5}+\frac{1}{5}{\rm i}, p_2=\frac{4}{5}-\frac{2}{5}{\rm i}, c^{(1)}_1=1+\frac{1}{2}{\rm i},c^{(1)}_2=\frac{5}{2}, \rho_1=\rho_2=\alpha_1=1, \alpha_2=\frac{1}{2}, \xi_{10}=\xi_{20}=0$ at time $t=0$ under the same nonlinear coefficients as previous.
In Fig.\ref{mix-fig5}, (a) and (d) represent the collisions of two bright solitons in the $\Phi^{(1)}$ and $u$ components, respectively; and the collisions of two dark solitons with different amplitudes in the $\Phi^{(2)}$ and $\Phi^{(3)}$ components are displayed in (b) and (c), respectively. From these figures, it is obvious that the solitons in all the components undergo elastic collision without shape change while just accompanied by a position shift.


\section{Bright-dark Mixed $N$-soliton Solution of the Multi-component Mel'nikov System}
In this section, we consider the general bright-dark mixed $N$-soliton solution consisting of $m$ bright solitons and $M-m$ dark
solitons in the short-wave components to the $M$-component Mel'nikov system (\ref{j3})-(\ref{j4}). To this end, the following dependent variable transformations are introduced
\begin{align}
& \Phi^{(k)}= \frac{g^{(k)}}{f},\ \ \ \ \Phi^{(l)}=\rho_l {\rm e}^{\textmd{i}\theta_l} \frac{h^{(l)}}{f},\ \ \ \ u=2 (\log f)_{xx},
\end{align}
where $\theta_l=\alpha_lx+\alpha^2_ly+\beta_l(t)$; $k=1,2,\cdots,m; l=1,2,\cdots,M-m$; which convert equations (\ref{j3})-(\ref{j4}) into
\begin{align}
& (D^2_x-{\rm i} D_y)g^{(k)} \cdot f=0,\ \ \ \ \ \ \ k=1,2,\cdots,m,\label{jj94}\\
& (D^2_x+2 \textmd{i} \alpha_l D_x-{\rm i} D_y)h^{(l)} \cdot f=0,\ \ \ \ \ \ \ l=1,2,\cdots,M-m,\label{jj95}\\
& (D^4_x+D_xD_t-3D^2_y)f\cdot f=-\sum^m_{k=1} \sigma_k g^{(k)}{g^{(k)}}^\ast+\sum^{M-m}_{l=1}\sigma_{l+m} \rho^2_l (f^2-h^{(l)}{h^{(l)}}^\ast).\label{jj96}
\end{align}

In the same manner as the three-component Mel'nikov system (\ref{j5})-(\ref{j8}), one can show that the following tau functions satisfy the bilinear equations (\ref{jj94})-(\ref{jj96}) and thus provide the general mixed $N$-soliton solution to the $M$-component Mel'nikov system (\ref{j3})-(\ref{j4})
\begin{align}\label{j105}
& f=\left| \begin{array}{ccccc}
A & I  \\
-I & B
\end{array} \right|,\ \ \ \ \ \ \
 g^{(k)}=\left| \begin{array}{ccccc}
A & I  & \Omega^{\textmd{T}}\\
-I & B & \mathbf{0}^{\textmd{T}}\\
\mathbf{0} & C_k & 0
\end{array} \right|,\ \ \ \ \ \ \
h^{(l)}=\left| \begin{array}{ccccc}
A^{(l)} & I  \\
-I & B
\end{array} \right|,
\end{align}
where $A,A^{(l)}$ and $B$ are $N \times N$ matrices whose entries are given respectively as
\begin{align*}
& a_{ij}=\frac{1}{p_i+p^*_j} \textmd{e}^{\xi_i+\xi^*_j},\ \ \ \ \ \  a^{(l)}_{ij}=\frac{1}{p_i+p^*_j} \Big(-\frac{p_i-\textmd{i}\alpha_l}{p^*_j+\textmd{i}\alpha_l}\Big) \textmd{e}^{\xi_i+\xi^*_j},\\
&b_{ij}=\Big(\sum^m_{k=1}\sigma_kc_i^{(k)*}c_j^{(k)}\Big)\Big[8({p_i^*}^3+p_j^3)+\sum^{M-m}_{l=1}\frac{\sigma_{l+m}\rho^2_l(p_i^*+p_j)}{(p_i^*+\textmd{i}\alpha_l)(p_j-\textmd{i}\alpha_l)}\Big]^{-1},
\end{align*}
meanwhile, $\Omega$ and $C_k$ are $N$-component row vectors
\begin{align*}
& \Omega=(\textmd{e}^{\xi_1},\textmd{e}^{\xi_2},\cdots,\textmd{e}^{\xi_N}),\ \ \ \ \ \ C_k=-(c^{(k)}_1,c^{(k)}_2,\cdots,c^{(k)}_N),
\end{align*}
with $\xi_i=p_ix -\textmd{i}p^2_iy-8p^3_i t  + \xi_{i0}$; $p_i$, $\xi_{i0}$ and $c^{(k)}_i$, $(i=1,2,\cdots,N)$ are complex constants.

Similarly, a necessary condition of the $M$-component Mel'nikov system (\ref{j3})-(\ref{j4}) for the existence of a nonsingular mixed $N$-soliton solution is given by
\begin{align}
& \Big(\sum^m_{k=1}\sigma_k|c_i^{(k)}| ^2\Big)\Big(8(p^2_{iR}-3p^2_{iI})+\sum^{M-m}_{l=1}\frac{\sigma_{l+m}\rho^2_l}{|p_i-\textmd{i}\alpha_l|^2}\Big)>0,\ \ \ \ \ \ i=1,2,\cdots,N.
\end{align}

The formula obtained admits bright-dark mixed $N$-soliton solution to the $M$-component Mel'nikov system (\ref{j3})-(\ref{j4}) with all possible combinations of nonlinearities, including all-positive, all-negative and mixed types.
Besides, as pointed in Ref.\cite{lak3}, the arbitrariness of nonlinearities $\sigma_k$ increases the freedom which results in rich soliton dynamics. In parallel with the vector NLS equation \cite{feng} and the multi-component YO system \cite{chen2}, the expression of the general bright-dark mixed $N$-soliton solution also includes the all-bright and all-dark $N$-soliton solutions as special cases. For instance, the $N$-bright soliton solution can be directly obtained from the formula (\ref{j105}) by taking $m=M$; hence, it supports the same determinant form as the mixed $N$-soliton solution. Whereas, the expression of the $N$-dark soliton solution is different from the one of the mixed $N$-soliton solution. As discussed in Refs.\cite{feng} and \cite{chen2}, it is known that the general $N$-dark soliton solution can alternatively take the same form as (\ref{j105}) except redefining the matrix $B$ to be an identity matrix ($b_{ij}=\delta_{ij}$) and imposing the following constraints on the parameters
\begin{align}
& 8({p_i^*}^2- |p_i|^2 +p_i^2)+\sum^{M}_{l=1}\frac{\sigma_{l}\rho^2_l}{|p_i-\textmd{i}\alpha_l|^2}=0,\ \ \ \ \ \ i=1,2,\cdots,N.
\end{align}

For the collision of two solitons, the similar asymptotic analysis can be performed as in the above section, whose details are omitted here. It can be concluded that for a $M$-component Mel'nikov system (\ref{j3})-(\ref{j4}) with $M \geq 3$, energy-exchanging inelastic collision is possible only if the bright parts of the mixed solitons appear at least in two short-wave components. What's more, the bright solitons in the short-wave components take elastic collision when $\frac{|c^{(1)}_1|}{|c^{(1)}_2|}=\frac{|c^{(2)}_1|}{|c^{(2)}_2|}=\cdots=\frac{|c^{(k)}_1|}{|c^{(k)}_2|},k=1,2,\cdots,m$. Otherwise, they take energy-exchanging inelastic collision characterized by an intensity redistribution.
Whereas, the dark solitons in the short-wave components and the bright solitons in the long-wave component always undergo elastic collision without shape change while just accompanied by a position shift.

\section{Conclusion}
The general bright-dark mixed $N$-soliton solution of the multi-component Mel'nikov system with all possible combinations of nonlinearities is constructed by virtue of the KP hierarchy reduction technique. Taking the three-component Mel'nikov system as a concrete example, its two kinds of mixed $N$-soliton solution (two-bright-one-dark soliton and one-bright-two-dark soliton in the three short-wave components) are derived from the tau functions of the KP hierarchy in detail.
It is worth noting that the derivation of two-bright-one-dark mixed $N$-soliton solution starts from a (2+1)-component KP hierarchy with one copy of shifted singular point ($c_1$).
In contrast, for the construction of one-bright-two-dark mixed $N$-soliton solution, we start from a (1+1)-component KP hierarchy with two copies of shifted singular points ($c_1$ and $c_2$).
Hence, it is not difficult to conclude that the number of components in the KP hierarchy matches the number of short-wave components supporting bright solitons
while the number of the copies of shifted singular points equals to the number of short-wave components supporting dark solitons.
This fact also can be referred to the constructions in the Ref.\cite{jimbo1}.
Then we extend our analysis to the $M$-component Mel'nikov system to obtain its $m$-bright-($M-m$)-dark mixed $N$-soliton solution.
It is obvious that the mixed soliton solution can be derived from the reduction of a $(m+1)$-component KP
hierarchy with $M-m$ copies of shifted singular points.
The formula obtained also includes the general all-bright and all-dark $N$-soliton solutions as special cases.

Furthermore, the dynamics of single and two solitons are also discussed in detail. Particularly, for the collision of two solitons, it has been shown that for a $M$-component Mel'nikov system with $M \geq 3$, if the bright solitons appear at least in two short-wave components, then interesting collision behaviors take place, resulting in energy exchange among the bright solitons in the short-wave components. Generally, after the inelastic collision of two solitons, the intensity of a soliton is enhanced while the intensity of the other soliton is suppressed, which can be observed in (a) and (b) of Fig.\ref{mix-fig2}. Whereas, the dark solitons appearing in the short-wave components and the bright solitons appearing in the long-wave component always undergo elastic collision without shape change while just accompanied by a position shift. This interesting feature is in parallel with the one-dimensional and two-dimensional multi-component YO system \cite{lak1,chen2}.

\section{Acknowledgment}
We would like to express our sincere thanks to S.Y. Lou, J.C. Chen and other members of our discussion group
for their valuable comments and suggestions. The project is supported by the Global Change Research Program
of China (No.2015CB953904), National Natural Science Foundation of China (No.11675054 and 11435005), and
Shanghai Collaborative Innovation Center of Trustworthy Software for Internet of Things (No. ZF1213).

\end{document}